\documentclass[fleqn,usenatbib]{mnras}

\usepackage{amssymb,amsmath,framed}

\usepackage{bm}
\expandafter\ifx\csname package@font\endcsname\relax\else
 \expandafter\expandafter
 \expandafter\usepackage
 \expandafter\expandafter
 \expandafter{\csname package@font\endcsname}
\fi
\usepackage[T1]{fontenc}

\DeclareRobustCommand{\VAN}[3]{#2}
\let\VANthebibliography\thebibliography
\def\thebibliography{\DeclareRobustCommand{\VAN}[3]{##3}\VANthebibliography}

\usepackage{graphicx}
\usepackage{bm}
\usepackage{color}
\usepackage{newtxtext,newtxmath}

\title[Beyond Mediocrity]{Beyond Mediocrity: How Common is Life?}

\author[Balbi and Lingam]{Amedeo Balbi$^{1}$\thanks{E-mail: balbia@roma2.infn.it} and Manasvi Lingam$^{2}$
\\
$^{1}$Dipartimento di Fisica, Universit\`a di Roma ``Tor Vergata", 00133 Roma, Italy \\
$^{2}$Department of Aerospace, Physics and Space Sciences, Florida Institute of Technology, Melbourne FL 32901, USA\\
}
\date{Accepted XXX. Received YYY; in original form ZZZ}

\pubyear{2022}

\begin{document}
\label{firstpage}
\pagerange{\pageref{firstpage}--\pageref{lastpage}}
\maketitle

\begin{abstract}
The probability that life spontaneously emerges in a suitable environment (abiogenesis) is one of the major unknowns in astrobiology. Assessing its value is impeded by the lack of an accepted theory for the origin of life, and is further complicated by the existence of selection biases. Appealing uncritically to some version of the ``Principle of Mediocrity'' -- namely, the supposed typicality of what transpired on Earth -- is problematic on empirical or logical grounds. In this paper, we adopt a Bayesian statistical approach to put on rigorous footing the inference of lower bounds for the probability of abiogenesis, based on current and future evidence. We demonstrate that the single datum that life has appeared at least once on Earth merely sets weak constraints on the minimal probability of abiogenesis. In fact, the {\it a priori} probability assigned to this event (viz., optimistic, pessimistic or agnostic prior) exerts the strongest influence on the final result. We also show that the existence of a large number of habitable worlds does not necessarily imply, by itself, a high probability that life should be common in the universe. Instead, as delineated before, the choice of prior, which is subject to uncertainty (i.e., admits multiple scenarios), strongly influences the likelihood of life being common. If habitable worlds are uncommon, for an agnostic prior, a deterministic scenario for the origin of life might be favoured over one where abiogenesis is a fluke event.
\end{abstract}




\section{Introduction}
One of the major unknowns in astrobiology is the probability of abiogenesis, to wit, the spontaneous appearance of life on a suitable object -- which we label as a ``planet'', even though it may encompass moons and small bodies -- having some appropriate physicochemical conditions; many statistical analyses have sprung up to address this issue \citep[e.g.,][]{ST12,SC16,BG20,DK20}. According to conservative estimates, the number of planets in the Milky Way ($N$) is at least of the same order of magnitude as the number of stars \citep{CKB12}, or equivalently $N \sim 10^{11}$. If we broaden the net to include worlds with subsurface oceans, this number might increase by as much as $2$-$3$ orders of magnitude \citep{LL19,SJM21}.

The number $N_H$ of habitable planets, where life could indeed appear and survive, is more difficult to assess: estimates have ranged from merely a few in models inspired by the ``Rare Earth'' hypothesis \citep{WaB00} to the upper limit tantamount to $N_H \sim N$ \citep{VD15}, but there is reasonable hope of tightening the bounds with future observations \citep{FAD18,NM19}. It is worth highlighting that the term ``habitable'' is endowed with some ambiguity, as it is employed in the astrobiological literature to exemplify disparate properties of planetary bodies, ranging from allowing the prolonged presence of liquid water to being conducive for the appearance of life \citep[cf.][]{LBC09,CBB16,CSP19,RH20,CSS22}. In this paper, we will use it in the specific sense of ``having the proper set of conditions for abiogenesis''.\footnote{
A recent paper has proposed a new term for this property, i.e.\ ``urable'' \citep{Deamer2022}. However, we have opted to follow the canonical usage and preserved the label ``habitable'', with the above mentioned proviso.} The fact that we currently do not possess definite knowledge of what makes a planet suitable for the appearance of life suggests that we can treat $N_H$ as a free parameter. The probability $p_L$ that life appears on any such planet, however, is unconstrained on empirical grounds, and essentially impossible to predict given the lack of an accepted theory for the origin of life at the current stage \citep{SIW17,FSMI}. The probability $p_L$ is essentially the same as the factor $f_L$ that appears in the famous Drake equation \citep{Drake65,ISCS66}.

The absence of reliable estimates on the value of $p_L$ translates into the impossibility of guessing the expected number of inhabited planets in our Galaxy, $N_L=p_L\ N_H$, even after presuming that $N_H$ is well-constrained. A common argument for the prevalence of life in the universe relies on citing the sheer magnitude of $N$ (and, perhaps, of $N_H$): however, this datum by itself does not imply a large $N_L$, as $p_L$ could be made arbitrarily small without violating any current knowledge. In fact, as reviewed in \citet{SDO18}, some estimates for $p_L$ are many orders of magnitude lower than the inverse of the total number of stars (or even elementary particles) in the observable universe; two such striking examples of extremely low values of $p_L$ can be found in \citet{EK07,EVK11} and \citet{TT20}.

Equally inconclusive is the unqualified adoption of some version of the ``Principle of Mediocrity'', often conflated with the ``Copernican Principle'', which is predicated on the basic postulate that our situation/status is not privileged or special \citep[e.g.,][]{CS94}. This seemingly rational principle, whose origins are arguably traceable to cosmology, is reviewed in several sources \citep[e.g.,][]{DD01,MMC12,CS14,BSSM}. As we shall elucidate later in this paper, through a Bayesian analysis, embracing this stance without due care is rendered problematic from a statistical standpoint, and can consequently lead to questionable conclusions.

In this paper, we aim to place on a rigorous footing the attempt of inferring the value of $p_L$ from the available (and forthcoming) evidence. Our way of framing the treatment and discussion is to invoke an agnostic stance, and to explore the spectrum of possibilities ensconced within two extreme situations: (1) the ``deterministic'' or ``necessity'' scenario, where the appearance of life has a very high probability of instantiation given the right conditions \citep{CS95,CDD95}, and (2) the diametrically opposite ``fluke'' or ``chance'' scenario where life has an extremely low chance of appearing, even with access to conducive environments \citep{Mon71,SCM03,EVK11}; it is evident that many possibilities would straddle these two extremes.

If and until future observations are able to unequivocally resolve the preceding dispute, the sole empirical information we can rely on, besides the total number of planets -- or, more broadly, possible sites for abiogenesis -- in the Milky Way ($N$), is the fact the life does exist on at least one planet, namely, the Earth. By adopting a Bayesian approach, we show that this datum can be used to derive lower bounds (at a given confidence level) on the potential value of $p_L$. However, we will show that these bounds crucially depend both on our prior presuppositions concerning the probability of abiogenesis, and on the independent expectation about the number of habitable worlds.

The paper is organised as follows. In Section \ref{SecFal}, we offer a critique of the ``Principle of Mediocrity'' building on earlier works. Next, we tackle the crux of the paper, the Bayesian analysis of $p_L$, in Section \ref{SecBayes}, with some technical calculations relegated to Appendix \ref{AppA}. Finally, we summarise our salient findings and implications in Section \ref{SecConc}.

\section{The Problems with Mediocrity}\label{SecFal}
The so-called ``Principle of Mediocrity'' is a commonly employed assumption in astrobiological studies. In short, this amounts to assuming that our situation is not privileged or special, and that it can be considered a random sample drawn from some suitable set. As sensible as this stipulation may seem \textit{prima facie}, it becomes questionable on closer inspection. A number of publications have delved into this subject and pointed out the limitations and/or flaws associated with applying this principle \textit{tout court} \citep{FC81,AK10,CS14,HA14,CB20,BCL23}. Of particular noteworthiness is the extensive critique of this principle from a philosophical standpoint by \citet{RM93}.

For example, we may ask what, precisely, should we regard as ``our situation'' referenced in the prior paragraph: is it the fact that we are living organisms, or that we are intelligent observers, or that we inhabit a planet with certain specific geochemical attributes, or that we function by exploiting select biochemical mechanisms, or a combination of these facets, or something else? Directly related to this question is another one -- namely, what is the exact set that our sample was drawn from? Is it the set of all possible intelligent observers, or the set of habitable planets, or what else? 

The situation is further complicated by the oft-invoked false dichotomy between ``mediocrity'' and ``specialness'' \citep{CHL14,ML21}. These terms cannot and ought not be applied in an unreflective fashion, as they may both adequately describe certain aspects of the phenomenon or object under examination \citep[e.g.,][]{CS14}. For example, the Earth is clearly just one of a large number of (terrestrial) planets in the Milky Way, but its physical attributes (e.g., size; distance from its star; atmospheric composition) are not necessarily representative of the norm. 

As an illustration, the Earth orbits an \emph{atypical} star (of G2V type), to wit, the Sun \citep{RLG08}. The atypical aspect arises because M-dwarfs are both more common and long-lived than G-type stars; this topic (especially the implications for habitability) has been explored in several publications \citep{LBS16,DW17,HKW18,ML18,LL18,MaLi,SBS21}. Likewise, the Earth might be atypical in hosting both landmasses and oceans on its surface, in contrast to either desiccated planets or water worlds that could be more prevalent in our Galaxy \citep{TI15,MLi19,KI22,HS23,LBM23}.

In fact, astronomical observations indicate that ``Earth-like'' planets, however loosely defined, comprise a small fraction of the total \citep{ZD21}. Therefore, the mediocrity principle may be true if it means that Earth is typical among all planets that are similar to Earth (in some sense), but this statement is nothing more than a tautology and it is practically of no use for making sensible inferences. Furthermore, as some of the physical attributes of planets as well as their host stars are anticipated to exert a profound influence on their habitability \citep{CBB16,ML19}, the very operation of ``picking at random'' becomes dubious \citep{RM93}, as it could be biased by vital observation-selection effects \citep{Bos02}. To put it another way, we can only inhabit a planet whose features are amenable to the existence of intelligent observers, no matter how unlikely this outcome might be.

While the Principle of Mediocrity continues to be a strong guiding principle in many astrophysical situations -- particularly in its more restricted and well-defined Copernican version \citep[e.g.,][]{PV17}, to wit, that the universe should look on average the same from any position and in any direction -- it has long been recognised that its application is more precarious when the possible presence of observers strongly depends on some physical prerequisites \citep{BC83}. One way around these difficulties is to adopt an additional presupposition known as the ``self-sampling assumption'' \citep{Bos02}, namely, that we should consider ourselves as a random sample from the set of all observers in our reference class. Yet, there is no obvious reference class that applies in general to our situation: each particular problem would require a careful explication of the reference class, to avoid incurring statistical fallacies. 

Unfortunately, the Principle of Mediocrity is frequently adopted uncritically in astrobiology, which may lead to unsubstantiated claims. One widespread example is the statement that life \textit{must} be common in the universe, as otherwise our situation would be rendered ``special''. A similar variant of the argument is encountered in the standard lore that, just because there are a multitude of planets in the universe, the chances that life is widespread \textit{must} likewise be large. It is feasible to understand why such arguments are not airtight in every instance, and are thus prone to fallacy. 

Consider, for example, the following situation: an urn contains a large number $N$ of well-mixed red and white balls, in unknown proportion. 
We are informed that a ball has been picked from the urn, and that its colour is white. From this single datum, it appears rational to infer that the number of white balls in the urn should be $N_w\sim N$. More rigorously, this relation follows from utilising the unbiased estimator of the success probability $p=k/n$ of a binomial distribution, with number of successes $k=1$ and number of trials $n=1$, and noting that, by definition, $p=N_w/N$. We should therefore conclude that white balls are ``typical'' in the urn and that there is ``nothing special'' in picking one such white ball.

However, suppose that the previous narrative is altered as follows: we are informed that we have been sleeping for an indeterminate amount of time, during which an unknown number of balls were drawn from the urn, and only when the first white ball was picked, we were woken up. In this scenario, our single datum is essentially useless for inferring $p$, as we can \textit{only} observe a white ball (otherwise we would have kept on sleeping), and we will certainly do so at some point, as long as there is at least one white ball in the urn (i.e., provided that $p\neq 0$). In actuality, contrary to the na\"{i}ve expectation, increasing $N$ does not enhance the likelihood that there are many white balls in the urn in this specific context, but only reduces the minimum value of $p$ that happens to be compatible with our observation, since the sufficient condition is basically $p\ge 1/N$. 

Therefore, in this narrative there is no way of assessing the ``specialness'' or ``typicality'' of white balls. The outcome of the observation, from our point of view, would be precisely the same for both small and large values of $N_w$ (viz., for both $p\sim 0$ and $p\sim 1$). The situation delineated in this paragraph and the previous one is analogous to the one we experience as inhabitants of the Earth. We have woken up (so to speak) and recognised that we live on a planet, and we also observed that there is a very large number of planets ($N$) in the universe. We know nothing about the probability $p_L$ that life appears on a planet, but we are certain that, as long as $p_L\ne 0$, we should find ourselves on a planet where life appeared; the latter reasoning can be placed on a mathematical (Bayesian) footing, as done in \citet{BC83} and \citet{SIW17}.

Adopting the self-sampling assumption does not seem particularly helpful on first sight, as there is some ambiguity on what should be the right reference class. To illustrate this aspect, let us contemplate the following thought experiment. Suppose, for simplicity, that we pick our species from the set of all intelligent observers dwelling on a planet in the Milky Way. This reference class would comprise $N_O$ elements, where $N_O$ is the number of intelligent species in the galaxy. As per the Principle of Mediocrity, we should be a typical sample from this class. However, the set may very well contain just one element (our species), in which case considering our situation ``typical'' would amount to a rather empty or uninformative statement. 

Alternatively, we may then opt to break down the problem by using nested reference classes, so that, for instance, $N_O=p_O N_L$, where $N_L$ is the expected number of inhabited planets in the galaxy, and $p_O$ is the probability of an intelligent species emerging on such a planet. Again, we can certainly presume that the Earth is drawn randomly from the set of all inhabited planets in the galaxy, but we have no idea about how many elements exist in such a set, so we have just moved our lack of knowledge one level below. And this drawback propagates in the same vein from one level to the next.

Hence, at first glimpse in astrobiology, it would seem as though there might be little hope of anchoring the Principle of Mediocrity to some objectively defined reference class. In fact, as we will show in the remainder of this paper, our assessment of the probability of life's emergence depends in a rather non-trivial way on the value of $N_H$, and cannot be easily disentangled from it without invoking some additional presupposition(s).

\section{The Probability of Life: a Bayesian Approach}\label{SecBayes}

\begin{figure*}
\includegraphics[width=\textwidth]{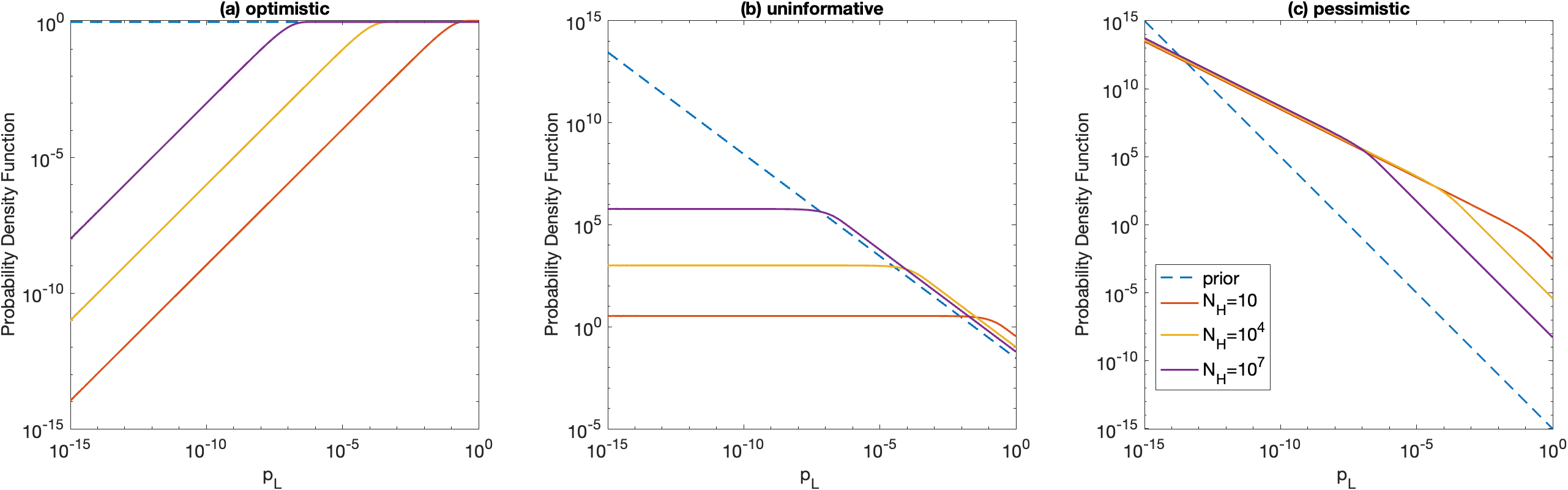}
\caption{The posterior probability distribution function (PDF) of $p_L$ (the probability of abiogenesis), given the evidence that life exists on Earth, computed from (\ref{eq:pdf}). The PDF is a function of the number of habitable worlds ($N_H$) in the Milky Way.}\label{fig:pdf}
\end{figure*}

\begin{figure*}
\includegraphics[width=\textwidth]{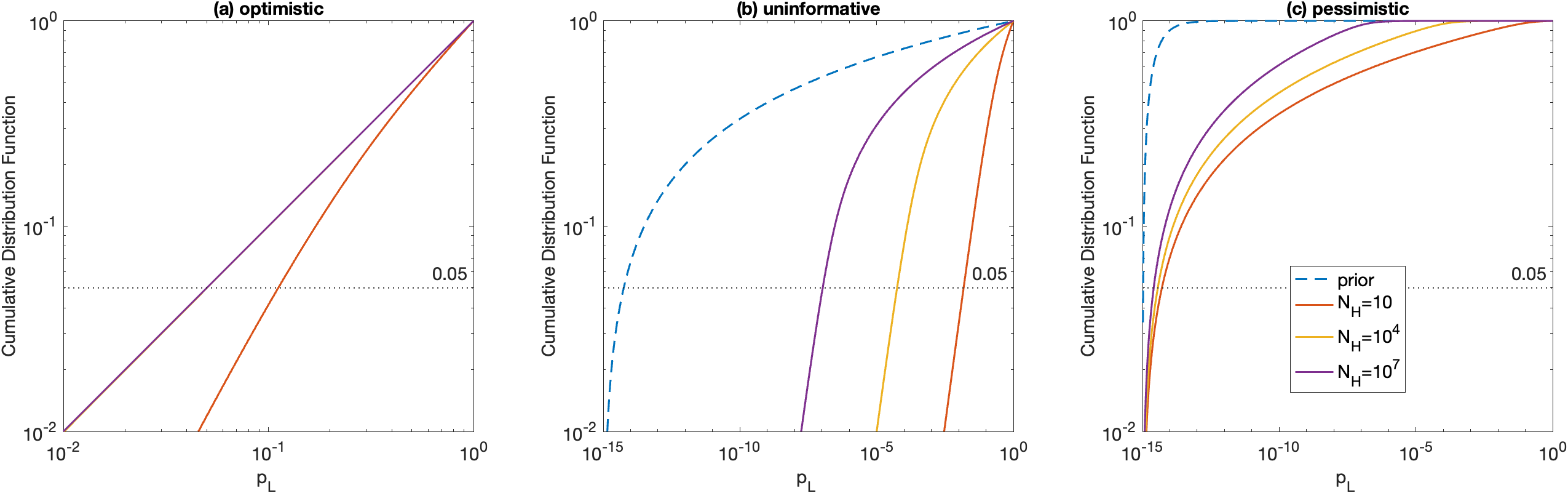}
\caption{The posterior cumulative distribution function (CDF) of $p_L$ (the probability of abiogenesis), given the evidence that life exists on Earth, computed from (\ref{eq:pdf}). The CDF is a function of the number of habitable worlds ($N_H$) in the Milky Way. For reference, the horizontal line represents a cumulative probability of $0.05$ and ties in directly with Table \ref{tab:pbounds}.}\label{fig:cdf}
\end{figure*}

In this section, using a Bayesian framework, we investigate how the number of habitable worlds ($N_H$) in the Milky Way, as well as crucial assumptions concerning the prior likelihood, regulate the bounds that can be placed on the probability of abiogenesis.

To make the discourse more definite, the expected number of inhabited planets is expressed as $N_L=p_L\ N_H$, where $p_L\in [0,1]$ signifies the unknown probability for the appearance of life given the right conditions, assumed for simplicity to possess a homogeneous spatiotemporal value.\footnote{In actuality, $p_L$ should vary from one habitable world to another, owing to which we may interpret this variable as some kind of spatiotemporal average.} We can then model the appearance of life in the galaxy as a stochastic process over repeated trials, with $p_L$ representing the probability of success, $\tilde{N}_L$ the number of successful trials, and $N_H$ the total number of trials. The probability function for such a process is a binomial distribution:
\begin{equation}\label{binomial}
P(\tilde{N}_L;N_H,p_L)= \binom{N_H}{\tilde{N}_L}p_L^{\tilde{N}_L}(1-p_L)^{N_H-\tilde{N}_L}
\end{equation}
How is $p_L$ potentially modulated by the magnitude of $N_H$? We can seek an answer to this question as described hereafter. We know that life appeared on at least one planet ($\tilde{N}_L\geq 1$); let us call this the evidence, $E$. The likelihood of this process within the model delineated above may be readily computed by first evaluating (\ref{binomial}) for the case of zero successful events ($\tilde{N}_L=0$):
\begin{equation}
P(0;N_H,p_L)= \binom{N_H}{0} p_L^{0}(1-p_L)^{N_H}=(1-p_L)^{N_H}
\end{equation}
and then noting that $P(\tilde{N}_L\geq 1;N_H,p_L)=1-P(0;N_H,p_L)$ by the rule of complementary events. Therefore, the likelihood of obtaining the aforementioned evidence $E$ is simply:
\begin{equation}
P(E\vert p_L)= 1 - (1-p_L)^{N_H}
\end{equation}

On applying Bayes theorem, we are in a position to calculate the posterior probability of $p_L$ given the evidence $E$ as:
\begin{equation}\label{eq:pdf}
P(p_L \vert E)= \frac{P(E \vert p_L) P(p_L)}{\int dp_L P(E \vert p_L) P(p_L)}
\end{equation}
where $P(p_L)$ is the prior probability of $p_L$, which is unknown as we lack a comprehensive theory for abiogenesis even on Earth, let alone on other worlds. Since we have no idea about the most likely value of $p_L$, we adopt three different priors reflecting different beliefs about the probability that life is manifested given the right conditions. 
\begin{itemize}
\item The prior $P(p_L) = \mathrm{constant}$, that is uniform in $p_L$, which therefore confers more weight upon values $p_L\sim 1$. This choice would represent an optimistic or deterministic scenario for abiogenesis.
\item The prior $P(p_L)\propto 1/p_L^2$, which is uniform in $p_L^{-1}$ and accords more weight to values $p_L\sim 0$, thereby corresponding to a pessimistic or fluke scenario for the origin of life. 
\item The log-uniform prior $P(p_L)\propto 1/p_L$ gives equal weight to all orders of magnitude of $p_L$, and may consequently be considered neither optimistic nor pessimistic; this prior is usually termed ``uninformative'' in statistical parlance.
\end{itemize}
By treating $N_H$ as an unknown free parameter that might be potentially constrained in the future, we can estimate the posterior probability of $p_L$ given the evidence. When numerically calculating the posterior $P(p_L \vert E)$, we will truncate the integration at a minimum value $p_{L, \rm min}=10^{-15}$ in order to avoid divergences at the unphysical value $p_L=0$. This value is much lower than the probability that life appeared on at least one planet in our Galaxy\footnote{This value is much higher than the probability that life emerged on a minimum of one world in the observable universe \citep{FS16}, but this would not change the overall logic of the argument.}. Analytical formulae enabling the calculation of the posterior are derived in Appendix \ref{AppA}.

\begin{table}
    \centering
    \begin{tabular}{l c c c}
          & optimistic & uninformative & pessimistic \\
         \hline
         prior & $4.9 \times 10^{-2}$ & $5.4 \times 10^{-15}$ & $1.0 \times 10^{-15}$ \\
          \hline
         post., $N_H=10$ & $1.1 \times 10^{-1}$ & $1.5 \times 10^{-2}$ & $5.1 \times 10^{-15}$\\ 
         post., $N_H=10^4$ & $4.9 \times 10^{-2}$ & $5.5 \times 10^{-5}$ & $3.6 \times 10^{-15}$\\
         post., $N_H=10^7$ & $4.9 \times 10^{-2}$ & $1.0 \times 10^{-7}$ & $2.5 \times 10^{-15}$\\
    \end{tabular}
    \caption{Lower bound on the probability of abiogenesis ($p_L$) at 95\% confidence level, assuming the cutoff $p_{L, \rm min}=10^{-15}$. The first row corresponds to the bounds derived from the optimistic, uninformative, and pessimistic priors, whereas the remaining rows show the bounds for the posteriors, where the evidence that life exists on Earth is taken into account. The bounds are a function of the number of habitable worlds ($N_H$) in the Milky Way.}
    \label{tab:pbounds}
\end{table}

Figure \ref{fig:pdf} shows the (posterior) probability density function (PDF) for $p_L$ computed from (\ref{eq:pdf}), after specifying three reference values for $N_H$ given by $N_H=10$, $10^4$ and $10^7$, ranging from pessimistic to optimistic. We have plotted Figure \ref{fig:pdf} as a log-log plot since both the posterior and $p_L$ span many orders of magnitude, thus rendering this representation well-suited. Although not immediately apparent, the integral of the PDF over all possible values of $p_L$ (for all choices of the prior and $N_H$) has been numerically verified to equal unity, as required. Along expected lines, not only is the most likely value of $p_L$ inherently sensitive to the inputted prior probability, but is also impacted by the magnitude of $N_H$. It is apparent from the figure that the PDF increases with $N_H$ for a fixed $p_L$. However, it is equally important to recognize that a larger value of $N_H$ does not imply, \textit{per se}, an increased number of inhabited worlds, because the evidence (viz., the existence of life on Earth) is compatible with a wide interval of $p_L$, and strongly depends on the choice of prior. 

The above features are evident on inspecting Figure \ref{fig:cdf}, where we have plotted the (posterior) cumulative distribution function (CDF) for $p_L$; note that the CDF quantifies the probability that $p_L$ is smaller than a given magnitude. The advantage of the CDF is that it can be harnessed to estimate Bayesian credible intervals for $p_L$. For example, if CDF happens to be $0.05$ for a certain value of $p_L$, this magnitude would indicate that $p_L$ is larger than that value with $95\%$ confidence. We utilise the CDF to construct Table \ref{tab:pbounds}, which depicts the lower bounds on $p_L$ at 95\% confidence, based on different priors and on the three reference values for $N_H$.

The first row of Table \ref{tab:pbounds} shows the lower bound derived from the prior distribution; in other words, this is the bound we would have ascribed to $p_L$ before taking the evidence into account. As it is apparent, an optimist would deem $p_L$ to be at least $\sim 5 \times 10^{-2}$, as opposed to a pessimist who would assign the minimum chance of $p_{L, \rm min}$ to the appearance of life. This trend arises because the pessimist would incline toward $p_{L, \rm min}$, and the optimist would loosely favour $1 - \left(\mathrm{confidence\,level}/100\right)$; the confidence level is expressed as a percentage, hence the normalisation by $100$. The other rows represent the updated lower bound on $p_L$ after the evidence is taken into account, assuming increasing values for $N_H$. 

A noticeable aspect of Table \ref{tab:pbounds} is that the largest shift from prior belief is manifested when only a small number of planets can support the appearance of life. In general, the existence of a large number of habitable planets suggests that smaller values of $p_L$ may be reconciled with the evidence. This somewhat counter-intuitive feature is actually as expected, because the data that life indeed appeared on at least one planet would imply a larger $p_L$ if we were to know that merely a few trials are accessible. However, the shift with respect to the prior $p_L$ is marginal in the extreme scenarios: both an optimist and a pessimist would not change their initial beliefs substantially, regardless of the precise value of $N_H$. 

In qualitative terms, they are adhering to a predetermined stance, and will not abandon it. In fact, a pessimist would be justified in maintaining their prior belief that $p_L\sim p_{L, \rm min}$, and in interpreting their observation (namely, being alive on planet Earth) as a classic selection effect. On the other hand, the uninformative prior receives a substantial update, which in turn strongly depends on $N_H$. If there are, say, only $10$ habitable planets in the galaxy, an agnostic would be led to conclude that $p_L$ is at least $1.5\times 10^{-2}$ with $95\%$ confidence, a significant departure from the assumed \emph{a priori} bound of $p_L\ge 5.4\times 10^{-15}$. On the other hand, if the higher $N_H=10^7$ is inputted, $p_L\ge 5.4\times 10^{-7}$ would suffice to explain the evidence. 

Finally, we comment on the nature of $p_{L, \rm min}$. Changing its value has virtually no effect on the optimistic and agnostic scenarios, whereas in the pessimistic case the inferred lower bound closely mirrors that of $p_{L, \rm min}$, and can therefore become arbitrarily small. The behaviour associated with the pessimistic case is consistent with what one would anticipate, because it distinctly favours smaller values of $p_L$ that are relatively close to $p_{L, \rm min}$.

\section{Discussion and Conclusions}\label{SecConc}
A widely presumed ``corollary'', in a manner of speaking, of the Principle of Mediocrity is that an increase in the number of potential sites for abiogenesis (i.e., habitable worlds) should automatically boost the chances of finding life. By drawing on this supposition as our motivating factor, we performed a Bayesian analysis of the probability of abiogenesis ($p_L$). Our work delineates the profound influence of \emph{a priori} assumptions, and underscores the necessity for a careful statistical treatment of inference from sparse data. 

We modelled the number of habitable ``planets'' ($N_H$) as a free parameter, since we do not have a complete understanding of the necessary and sufficient conditions for a world to be deemed habitable, at both a particular instant in time and over an extended interval \citep{CBB16,ML21}. The word ``habitability'' is itself subject to some dispute, for example, as to whether it alludes to a binary or continuous property \citep[cf.][]{CSP19,RH20,CSS22}. Furthermore, even if we resolve the theoretical issues confronting habitability, we have not currently placed empirical constraints on $N_H$, thereby bolstering the case for envisaging and modelling $N_H$ as a free parameter.

Contrary to the aforementioned corollary, we demonstrated that the number of habitable worlds has an equivocal effect on the expected number of inhabited worlds ($N_L=p_L\ N_H$) in the Galaxy, and that the prior would have a more profound impact on $N_L$. To be specific, there are two key findings that are discernible from our analysis.
\begin{enumerate}
    \item Varying $N_H$ ostensibly has a minimal effect on $N_L$ for both the optimist and pessimist; the lower bound for $p_L$ is mostly independent of $N_H$, as seen from Table \ref{tab:pbounds}. 
    \item Varying $N_H$ does indeed exert some influence on the agnostic, but it roughly goes in the opposite direction with respect to the na\"{i}ve surmise that large values of $N_H$ are tantamount to a higher likelihood of extraterrestrial life. More precisely, as indicated by Table \ref{tab:pbounds}, increasing $N_H$ makes a smaller value of $p_L$ more compatible, but it does not directly shed light on the frequency of life elsewhere.
\end{enumerate}
Our work has a number of implications for exoplanet surveys and astrobiology in general, some of which are discussed below.

First and foremost, our analysis reinforces the centrality of the prior and its ability to ``wash out'' the evidence in certain cases. While this feature was recognised and underscored before -- notably in \citet{BC83}, \citet{ST12}, \citet{VH17}, \citet{BG20}, \citet{CHL22}, and \citet{LHW23}, among other publications -- the significance of the prior in this context (involving $p_L$ and $N_H$) has not been articulated before. Hence, our paper serves to highlight the crucial notion that, until a rigorous theory of living systems and the prior probability distribution $P(p_L)$ is available, we will not be able to accurately judge the frequency of abiogenesis and the number of inhabited worlds in the universe (see also \citealt{FSMI}).

Second, this treatment separates the biological and planetary horns of the fundamental question: what is the expected number of inhabited worlds in the Milky Way? We will focus on the agnostic scenario in Table \ref{tab:pbounds}, which is predicated on the uninformative prior, as it does not incline exclusively toward minuscule or relatively high values of $p_L$ unlike the optimistic and pessimistic cases.
\begin{itemize}
    \item The importance of developing frameworks to quantify $N_H$ is obvious, since this parameter modulates the plausible lower bound on $p_L$. If future studies (both theoretical and observational) are able to constrain $N_H$, we may take advantage of the Bayesian approach (Table \ref{tab:pbounds} in particular) and obtain the potential lower bound for $p_L$, perhaps even if no exoplanetary biosignatures are detected.
    \item On account of the approximate inverse correlation between $N_H$ and the lower bound for $p_L$ evinced by Table \ref{tab:pbounds}, it is conceivable that ``Rare Earth'' hypotheses might favour the emergence of life on habitable worlds. The reason is that such hypotheses suggest that truly habitable worlds (akin to Earth) are rare because a number of criteria (e.g., large moon) must be met. A low value of $N_H$ would, in turn, be compatible with an enhanced lower bound for $p_L$. 
    \item Continuing with the above narrative, if $p_L$ is on the higher side (and $N_H$ is commensurately low), this regime apparently favours the deterministic picture wherein the origin of life is rendered probable in a habitable environment, albeit with the caveat that such habitable settings are anticipated to be rare. The converse situation is tenable when $N_H$ is elevated to a high value.
    \item As per the explicit and implicit postulates of our formalism, we see that the product of $N_H$ and the lower bound on $p_L$ is around unity for the agnostic case; this relation is essentially the quantitative version of the statement that we would tend to forecast $N_L \geq 1$ in our Galaxy (by virtue of our own existence).
\end{itemize}

In closing, we point out that $N_H$ was treated as an input parameter, and we derived lower bounds (at a certain confidence level) on $p_L$. However, this splitting was motivated purely by the standard expectation that it is easier to pin down the magnitude or range of $N_H$ compared to $p_L$ in the future, owing to which we sought to constrain the lower bound of the latter. Instead, if we gain a deeper quantitative understanding of $p_L$ first, we may adopt it as the input parameter and thereby derive a lower bound on $N_H$. This line of enquiry could be pursued in the future, as it would complement our treatment.

\section*{Data Availability Statement}
No new data were generated or analysed in support of this research.

\bibliographystyle{mnras}
\bibliography{Mediocrity}

\begin{thebibliography}{}
\makeatletter
\relax
\def\mn@urlcharsother{\let\do\@makeother \do\$\do\&\do\#\do\^\do\_\do\%\do\~}
\def\mn@doi{\begingroup\mn@urlcharsother \@ifnextchar [ {\mn@doi@}
  {\mn@doi@[]}}
\def\mn@doi@[#1]#2{\def\@tempa{#1}\ifx\@tempa\@empty \href
  {http://dx.doi.org/#2} {doi:#2}\else \href {http://dx.doi.org/#2} {#1}\fi
  \endgroup}
\def\mn@eprint#1#2{\mn@eprint@#1:#2::\@nil}
\def\mn@eprint@arXiv#1{\href {http://arxiv.org/abs/#1} {{\tt arXiv:#1}}}
\def\mn@eprint@dblp#1{\href {http://dblp.uni-trier.de/rec/bibtex/#1.xml}
  {dblp:#1}}
\def\mn@eprint@#1:#2:#3:#4\@nil{\def\@tempa {#1}\def\@tempb {#2}\def\@tempc
  {#3}\ifx \@tempc \@empty \let \@tempc \@tempb \let \@tempb \@tempa \fi \ifx
  \@tempb \@empty \def\@tempb {arXiv}\fi \@ifundefined
  {mn@eprint@\@tempb}{\@tempb:\@tempc}{\expandafter \expandafter \csname
  mn@eprint@\@tempb\endcsname \expandafter{\@tempc}}}

\bibitem[\protect\citeauthoryear{{Abramowitz} \& {Stegun}}{{Abramowitz} \&
  {Stegun}}{1965}]{AbSte65}
{Abramowitz} M.,  {Stegun} I.~A.,  1965, {Handbook of Mathematical Functions
  with Formulas, Graphs, and Mathematical Tables}.
New York: Dover Publications, Inc.

\bibitem[\protect\citeauthoryear{{Balbi} \& {Grimaldi}}{{Balbi} \&
  {Grimaldi}}{2020}]{BG20}
{Balbi} A.,  {Grimaldi} C.,  2020, \mn@doi [Proc. Natl. Acad. Sci.]
  {10.1073/pnas.2007560117}, \href
  {https://ui.adsabs.harvard.edu/abs/2020PNAS..11721031B} {117, 21031}

\bibitem[\protect\citeauthoryear{{Bennett}, {Shostak}, {Schneider}  \&
  {MacGregor}}{{Bennett} et~al.}{2022}]{BSSM}
{Bennett} J.,  {Shostak} S.,  {Schneider} N.,   {MacGregor} M.,  2022, Life in
  the Universe, 5th edn.
Princeton: Princeton University Press

\bibitem[\protect\citeauthoryear{{Bostrom}}{{Bostrom}}{2002}]{Bos02}
{Bostrom} N.,  2002, {Anthropic Bias: Observation Selection Effects in Science
  and Philosophy}.
London: Routledge

\bibitem[\protect\citeauthoryear{{Carter}}{{Carter}}{1983}]{BC83}
{Carter} B.,  1983, \mn@doi [Phil. Trans. R. Soc. Lond. A]
  {10.1098/rsta.1983.0096}, \href
  {https://ui.adsabs.harvard.edu/abs/1983RSPTA.310..347C} {310, 347}

\bibitem[\protect\citeauthoryear{{Cassan} et~al.,}{{Cassan}
  et~al.}{2012}]{CKB12}
{Cassan} A.,  et~al., 2012, \mn@doi [Nature] {10.1038/nature10684}, \href
  {https://ui.adsabs.harvard.edu/abs/2012Natur.481..167C} {481, 167}

\bibitem[\protect\citeauthoryear{{{\'C}irkovi{\'c}}}{{{\'C}irkovi{\'c}}}{2012}]{MMC12}
{{\'C}irkovi{\'c}} M.~M.,  2012, {The Astrobiological Landscape: Philosophical
  Foundations of the Study of Cosmic Life}.
Cambridge: Cambridge University Press

\bibitem[\protect\citeauthoryear{{{\'C}irkovi{\'c}} \&
  {Balbi}}{{{\'C}irkovi{\'c}} \& {Balbi}}{2020}]{CB20}
{{\'C}irkovi{\'c}} M.~M.,  {Balbi} A.,  2020, \mn@doi [Int. J. Astrobiol.]
  {10.1017/S1473550419000223}, \href
  {https://ui.adsabs.harvard.edu/abs/2020IJAsB..19..101C} {19, 101}

\bibitem[\protect\citeauthoryear{{Cockell} et~al.,}{{Cockell}
  et~al.}{2016}]{CBB16}
{Cockell} C.~S.,  et~al., 2016, \mn@doi [Astrobiology] {10.1089/ast.2015.1295},
  \href {https://ui.adsabs.harvard.edu/abs/2016AsBio..16...89C} {16, 89}

\bibitem[\protect\citeauthoryear{{Cockell}, {Stevens}  \& {Prescott}}{{Cockell}
  et~al.}{2019}]{CSP19}
{Cockell} C.~S.,  {Stevens} A.~H.,   {Prescott} R.,  2019, \mn@doi [Nat.
  Astron.] {10.1038/s41550-019-0916-7}, \href
  {https://ui.adsabs.harvard.edu/abs/2019NatAs...3..956C} {3, 956}

\bibitem[\protect\citeauthoryear{{Cockell}, {Samuels}  \& {Stevens}}{{Cockell}
  et~al.}{2022}]{CSS22}
{Cockell} C.~S.,  {Samuels} T.,   {Stevens} A.~H.,  2022, \mn@doi
  [Astrobiology] {10.1089/ast.2021.0038}, \href
  {https://ui.adsabs.harvard.edu/abs/2022AsBio..22....7C} {22, 7}

\bibitem[\protect\citeauthoryear{{Crick}}{{Crick}}{1981}]{FC81}
{Crick} F.,  1981, {Life itself: its origin and nature.}.
New York: Simon and Schuster

\bibitem[\protect\citeauthoryear{{Darling}}{{Darling}}{2001}]{DD01}
{Darling} D.,  2001, {Life Everywhere: The Maverick Science of Astrobiology}.
New York: Basic Books

\bibitem[\protect\citeauthoryear{{De Duve}}{{De Duve}}{1995}]{CDD95}
{De Duve} C.,  1995, Vital Dust: Life As A Cosmic Imparative.
New York: Basic Books

\bibitem[\protect\citeauthoryear{{Deamer}, {Cary}  \& {Damer}}{{Deamer}
  et~al.}{2022}]{Deamer2022}
{Deamer} D.,  {Cary} F.,   {Damer} B.,  2022, \mn@doi [Astrobiology]
  {10.1089/ast.2021.0173}, \href
  {https://ui.adsabs.harvard.edu/abs/2022AsBio..22..889D} {22, 889}

\bibitem[\protect\citeauthoryear{{Drake}}{{Drake}}{1965}]{Drake65}
{Drake} F.~D.,  1965, {The Radio Search for Intelligent Extraterrestrial Life}.
Oxford: Pergamon Press, pp 323--345

\bibitem[\protect\citeauthoryear{{Foote}, {Sinhadc}, {Mathis}  \& {Imari
  Walker}}{{Foote} et~al.}{2022}]{FSMI}
{Foote} S.,  {Sinhadc} P.,  {Mathis} C.,   {Imari Walker} S.,  2022, arXiv
  e-prints, \href {https://ui.adsabs.harvard.edu/abs/2022arXiv220700634F} {p.
  arXiv:2207.00634}

\bibitem[\protect\citeauthoryear{{Frank} \& {Sullivan}}{{Frank} \&
  {Sullivan}}{2016}]{FS16}
{Frank} A.,  {Sullivan} W.~T. I.,  2016, \mn@doi [Astrobiology]
  {10.1089/ast.2015.1418}, \href
  {https://ui.adsabs.harvard.edu/abs/2016AsBio..16..359F} {16, 359}

\bibitem[\protect\citeauthoryear{{Fujii} et~al.,}{{Fujii} et~al.}{2018}]{FAD18}
{Fujii} Y.,  et~al., 2018, \mn@doi [Astrobiology] {10.1089/ast.2017.1733},
  \href {https://ui.adsabs.harvard.edu/abs/2018AsBio..18..739F} {18, 739}

\bibitem[\protect\citeauthoryear{{Haqq-Misra}, {Kopparapu}  \&
  {Wolf}}{{Haqq-Misra} et~al.}{2018}]{HKW18}
{Haqq-Misra} J.,  {Kopparapu} R.~K.,   {Wolf} E.~T.,  2018, \mn@doi [Int. J.
  Astrobiol.] {10.1017/S1473550417000118}, \href
  {https://ui.adsabs.harvard.edu/abs/2018IJAsB..17...77H} {17, 77}

\bibitem[\protect\citeauthoryear{{Heller}}{{Heller}}{2020}]{RH20}
{Heller} R.,  2020, \mn@doi [Nat. Astron.] {10.1038/s41550-020-1063-x}, \href
  {https://ui.adsabs.harvard.edu/abs/2020NatAs...4..294H} {4, 294}

\bibitem[\protect\citeauthoryear{{Heller} \& {Armstrong}}{{Heller} \&
  {Armstrong}}{2014}]{HA14}
{Heller} R.,  {Armstrong} J.,  2014, \mn@doi [Astrobiology]
  {10.1089/ast.2013.1088}, \href
  {https://ui.adsabs.harvard.edu/abs/2014AsBio..14...50H} {14, 50}

\bibitem[\protect\citeauthoryear{{H{\"o}ning} \& {Spohn}}{{H{\"o}ning} \&
  {Spohn}}{2023}]{HS23}
{H{\"o}ning} D.,  {Spohn} T.,  2023, \mn@doi [Astrobiology]
  {10.1089/ast.2022.0070}, \href
  {https://ui.adsabs.harvard.edu/abs/2023AsBio..23..372H} {23, 372}

\bibitem[\protect\citeauthoryear{{Imari Walker}}{{Imari Walker}}{2017}]{SIW17}
{Imari Walker} S.,  2017, \mn@doi [Rep. Prog. Phys.]
  {10.1088/1361-6633/aa7804}, \href
  {https://ui.adsabs.harvard.edu/abs/2017RPPh...80i2601I} {80, 092601}

\bibitem[\protect\citeauthoryear{{Kimura} \& {Ikoma}}{{Kimura} \&
  {Ikoma}}{2022}]{KI22}
{Kimura} T.,  {Ikoma} M.,  2022, \mn@doi [Nat. Astron.]
  {10.1038/s41550-022-01781-1}, \href
  {https://ui.adsabs.harvard.edu/abs/2022NatAs...6.1296K} {6, 1296}

\bibitem[\protect\citeauthoryear{{Kipping}}{{Kipping}}{2020}]{DK20}
{Kipping} D.,  2020, \mn@doi [Proc. Natl. Acad. Sci.]
  {10.1073/pnas.1921655117}, \href
  {https://ui.adsabs.harvard.edu/abs/2020PNAS..11711995K} {117, 11995}

\bibitem[\protect\citeauthoryear{{Koonin}}{{Koonin}}{2007}]{EK07}
{Koonin} E.~V.,  2007, \mn@doi [Biol. Direct] {10.1186/1745-6150-2-15}, 2, 1

\bibitem[\protect\citeauthoryear{{Koonin}}{{Koonin}}{2011}]{EVK11}
{Koonin} E.~V.,  2011, {The Logic of Chance: The Nature and Origin of
  Biological Evolution}.
Upper Saddle River: FT Press Science

\bibitem[\protect\citeauthoryear{{Kukla}}{{Kukla}}{2010}]{AK10}
{Kukla} A.,  2010, {Extraterrestrials: A Philosophical Perspective}.
Lanham: Lexington Books

\bibitem[\protect\citeauthoryear{{Lacki}}{{Lacki}}{2023}]{BCL23}
{Lacki} B.~C.,  2023, \mn@doi [Int. J. Astrobiol.] {10.1017/S1473550423000071},
  \href {https://ui.adsabs.harvard.edu/abs/2021arXiv210607738L} {pp 1--45}

\bibitem[\protect\citeauthoryear{{Lammer} et~al.,}{{Lammer}
  et~al.}{2009}]{LBC09}
{Lammer} H.,  et~al., 2009, \mn@doi [Astron. Astrophys. Rev.]
  {10.1007/s00159-009-0019-z}, \href
  {https://ui.adsabs.harvard.edu/abs/2009A&ARv..17..181L} {17, 181}

\bibitem[\protect\citeauthoryear{{Lineweaver}}{{Lineweaver}}{2014}]{CHL14}
{Lineweaver} C.~H.,  2014, \mn@doi [Orig. Life Evol. Biosph.]
  {10.1007/s11084-014-9369-2}, \href
  {https://ui.adsabs.harvard.edu/abs/2014OLEB...44..159L} {44, 159}

\bibitem[\protect\citeauthoryear{{Lineweaver}}{{Lineweaver}}{2022}]{CHL22}
{Lineweaver} C.~H.,  2022, \mn@doi [Astrobiology] {10.1089/ast.2021.0185},
  \href {https://ui.adsabs.harvard.edu/abs/2022AsBio..22.1419L} {22, 1419}

\bibitem[\protect\citeauthoryear{{Lingam} \& {Loeb}}{{Lingam} \&
  {Loeb}}{2018a}]{ML18}
{Lingam} M.,  {Loeb} A.,  2018a, \mn@doi [Astrophys. J. Lett.]
  {10.3847/2041-8213/aabd86}, \href
  {https://ui.adsabs.harvard.edu/abs/2018ApJ...857L..17L} {857, L17}

\bibitem[\protect\citeauthoryear{{Lingam} \& {Loeb}}{{Lingam} \&
  {Loeb}}{2018b}]{LL18}
{Lingam} M.,  {Loeb} A.,  2018b, \mn@doi [J. Cosmol. Astropart. Phys.]
  {10.1088/1475-7516/2018/05/020}, \href
  {https://ui.adsabs.harvard.edu/abs/2018JCAP...05..020L} {2018, 020}

\bibitem[\protect\citeauthoryear{{Lingam} \& {Loeb}}{{Lingam} \&
  {Loeb}}{2019a}]{LL19}
{Lingam} M.,  {Loeb} A.,  2019a, \mn@doi [Int. J. Astrobiol.]
  {10.1017/S1473550418000083}, \href
  {https://ui.adsabs.harvard.edu/abs/2019IJAsB..18..112L} {18, 112}

\bibitem[\protect\citeauthoryear{{Lingam} \& {Loeb}}{{Lingam} \&
  {Loeb}}{2019b}]{MaLi}
{Lingam} M.,  {Loeb} A.,  2019b, \mn@doi [Int. J. Astrobiol.]
  {10.1017/S1473550419000016}, \href
  {https://ui.adsabs.harvard.edu/abs/2019IJAsB..18..527L} {18, 527}

\bibitem[\protect\citeauthoryear{{Lingam} \& {Loeb}}{{Lingam} \&
  {Loeb}}{2019c}]{ML19}
{Lingam} M.,  {Loeb} A.,  2019c, \mn@doi [Rev. Mod. Phys.]
  {10.1103/RevModPhys.91.021002}, \href
  {https://ui.adsabs.harvard.edu/abs/2019RvMP...91b1002L} {91, 021002}

\bibitem[\protect\citeauthoryear{{Lingam} \& {Loeb}}{{Lingam} \&
  {Loeb}}{2019d}]{MLi19}
{Lingam} M.,  {Loeb} A.,  2019d, \mn@doi [Astron. J.]
  {10.3847/1538-3881/aaf420}, \href
  {https://ui.adsabs.harvard.edu/abs/2019AJ....157...25L} {157, 25}

\bibitem[\protect\citeauthoryear{{Lingam} \& {Loeb}}{{Lingam} \&
  {Loeb}}{2021}]{ML21}
{Lingam} M.,  {Loeb} A.,  2021, {Life in the Cosmos: From Biosignatures to
  Technosignatures}.
Cambridge: Harvard University Press

\bibitem[\protect\citeauthoryear{{Lingam}, {Haqq-Misra}, {Wright}, {Huston},
  {Frank}  \& {Kopparapu}}{{Lingam} et~al.}{2023a}]{LHW23}
{Lingam} M.,  {Haqq-Misra} J.,  {Wright} J.~T.,  {Huston} M.~J.,  {Frank} A.,
  {Kopparapu} R.,  2023a, \mn@doi [Astrophys. J.] {10.3847/1538-4357/acaca0},
  \href {https://ui.adsabs.harvard.edu/abs/2023ApJ...943...27L} {943, 27}

\bibitem[\protect\citeauthoryear{{Lingam}, {Balbi}  \& {Mahajan}}{{Lingam}
  et~al.}{2023b}]{LBM23}
{Lingam} M.,  {Balbi} A.,   {Mahajan} S.~M.,  2023b, \mn@doi [Astrophys. J.]
  {10.3847/1538-4357/acb6fa}, \href
  {https://ui.adsabs.harvard.edu/abs/2023ApJ...945...23L} {945, 23}

\bibitem[\protect\citeauthoryear{{Loeb}, {Batista}  \& {Sloan}}{{Loeb}
  et~al.}{2016}]{LBS16}
{Loeb} A.,  {Batista} R.~A.,   {Sloan} D.,  2016, \mn@doi [J. Cosmol.
  Astropart. Phys.] {10.1088/1475-7516/2016/08/040}, \href
  {https://ui.adsabs.harvard.edu/abs/2016JCAP...08..040L} {2016, 040}

\bibitem[\protect\citeauthoryear{{Madhusudhan}}{{Madhusudhan}}{2019}]{NM19}
{Madhusudhan} N.,  2019, \mn@doi [Annu. Rev. Astron. Astrophys.]
  {10.1146/annurev-astro-081817-051846}, \href
  {https://ui.adsabs.harvard.edu/abs/2019ARA&A..57..617M} {57, 617}

\bibitem[\protect\citeauthoryear{{Mash}}{{Mash}}{1993}]{RM93}
{Mash} R.,  1993, \mn@doi [Philos. Sci.] {10.1086/289729}, 60, 204

\bibitem[\protect\citeauthoryear{{Mojzsis}}{{Mojzsis}}{2021}]{SJM21}
{Mojzsis} S.~J.,  2021, \mn@doi [Nat. Astron.] {10.1038/s41550-021-01529-3},
  \href {https://ui.adsabs.harvard.edu/abs/2021NatAs...5.1083M} {5, 1083}

\bibitem[\protect\citeauthoryear{{Monod}}{{Monod}}{1971}]{Mon71}
{Monod} J.,  1971, {Chance and Necessity: An Essay on the Natural Philosophy of
  Modern Biology}.
New York: Alfred A. Knopf

\bibitem[\protect\citeauthoryear{{Morris}}{{Morris}}{2003}]{SCM03}
{Morris} S.~C.,  2003, {Life's Solution: Inevitable Humans in a Lonely
  Universe}.
Cambridge: Cambridge University Press

\bibitem[\protect\citeauthoryear{{Perlov} \& {Vilenkin}}{{Perlov} \&
  {Vilenkin}}{2017}]{PV17}
{Perlov} D.,  {Vilenkin} A.,  2017, {Cosmology for the Curious}.
Cham: Springer, \mn@doi{10.1007/978-3-319-57040-2}

\bibitem[\protect\citeauthoryear{{Robles}, {Lineweaver}, {Grether}, {Flynn},
  {Egan}, {Pracy}, {Holmberg}  \& {Gardner}}{{Robles} et~al.}{2008}]{RLG08}
{Robles} J.~A.,  {Lineweaver} C.~H.,  {Grether} D.,  {Flynn} C.,  {Egan} C.~A.,
   {Pracy} M.~B.,  {Holmberg} J.,   {Gardner} E.,  2008, \mn@doi [Astrophys.
  J.] {10.1086/589985}, \href
  {https://ui.adsabs.harvard.edu/abs/2008ApJ...684..691R} {684, 691}

\bibitem[\protect\citeauthoryear{{Sagan}}{{Sagan}}{1994}]{CS94}
{Sagan} C.,  1994, {Pale Blue Dot: A Vision of the Human Future in Space}.
New York: Random House

\bibitem[\protect\citeauthoryear{{Sagan}}{{Sagan}}{1995}]{CS95}
{Sagan} C.,  1995, Bioastron. News, 7, 1

\bibitem[\protect\citeauthoryear{{Sandberg}, {Drexler}  \& {Ord}}{{Sandberg}
  et~al.}{2018}]{SDO18}
{Sandberg} A.,  {Drexler} E.,   {Ord} T.,  2018, arXiv e-prints, \href
  {https://ui.adsabs.harvard.edu/abs/2018arXiv180602404S} {p. arXiv:1806.02404}

\bibitem[\protect\citeauthoryear{{Scharf}}{{Scharf}}{2014}]{CS14}
{Scharf} C.,  2014, {The Copernicus Complex: Our Cosmic Significance in a
  Universe of Planets and Probabilities}.
New York: Scientific American/Farrar, Straus and Giroux

\bibitem[\protect\citeauthoryear{{Scharf} \& {Cronin}}{{Scharf} \&
  {Cronin}}{2016}]{SC16}
{Scharf} C.,  {Cronin} L.,  2016, \mn@doi [Proc. Natl. Acad. Sci.]
  {10.1073/pnas.1523233113}, \href
  {https://ui.adsabs.harvard.edu/abs/2016PNAS..113.8127S} {113, 8127}

\bibitem[\protect\citeauthoryear{{Shklovskii} \& {Sagan}}{{Shklovskii} \&
  {Sagan}}{1966}]{ISCS66}
{Shklovskii} I.~S.,  {Sagan} C.,  1966, {Intelligent life in the universe}.
Holden-Day: San Francisco

\bibitem[\protect\citeauthoryear{{Snyder-Beattie}, {Sandberg}, {Drexler}  \&
  {Bonsall}}{{Snyder-Beattie} et~al.}{2021}]{SBS21}
{Snyder-Beattie} A.~E.,  {Sandberg} A.,  {Drexler} K.~E.,   {Bonsall} M.~B.,
  2021, \mn@doi [Astrobiology] {10.1089/ast.2019.2149}, \href
  {https://ui.adsabs.harvard.edu/abs/2021AsBio..21..265S} {21, 265}

\bibitem[\protect\citeauthoryear{{Spiegel} \& {Turner}}{{Spiegel} \&
  {Turner}}{2012}]{ST12}
{Spiegel} D.~S.,  {Turner} E.~L.,  2012, \mn@doi [Proc. Natl. Acad. Sci.]
  {10.1073/pnas.1111694108}, \href
  {https://ui.adsabs.harvard.edu/abs/2012PNAS..109..395S} {109, 395}

\bibitem[\protect\citeauthoryear{{Tian} \& {Ida}}{{Tian} \& {Ida}}{2015}]{TI15}
{Tian} F.,  {Ida} S.,  2015, \mn@doi [Nat. Geosci.] {10.1038/ngeo2372}, \href
  {https://ui.adsabs.harvard.edu/abs/2015NatGe...8..177T} {8, 177}

\bibitem[\protect\citeauthoryear{{Totani}}{{Totani}}{2020}]{TT20}
{Totani} T.,  2020, \mn@doi [Sci. Rep.] {10.1038/s41598-020-58060-0}, \href
  {https://ui.adsabs.harvard.edu/abs/2020NatSR..10.1671T} {10, 1671}

\bibitem[\protect\citeauthoryear{{Vakoch} \& {Dowd}}{{Vakoch} \&
  {Dowd}}{2015}]{VD15}
{Vakoch} D.~A.,  {Dowd} M.~F.,  eds, 2015, {The Drake Equation}.
Cambridge: Cambridge University Press

\bibitem[\protect\citeauthoryear{{Verendel} \& {H{\"a}ggstr{\"o}m}}{{Verendel}
  \& {H{\"a}ggstr{\"o}m}}{2017}]{VH17}
{Verendel} V.,  {H{\"a}ggstr{\"o}m} O.,  2017, \mn@doi [Int. J. Astrobiol.]
  {10.1017/S1473550415000452}, \href
  {https://ui.adsabs.harvard.edu/abs/2017IJAsB..16...14V} {16, 14}

\bibitem[\protect\citeauthoryear{{Waltham}}{{Waltham}}{2017}]{DW17}
{Waltham} D.,  2017, \mn@doi [Astrobiology] {10.1089/ast.2016.1518}, \href
  {https://ui.adsabs.harvard.edu/abs/2017AsBio..17...61W} {17, 61}

\bibitem[\protect\citeauthoryear{{Ward} \& {Brownlee}}{{Ward} \&
  {Brownlee}}{2000}]{WaB00}
{Ward} P.,  {Brownlee} D.,  2000, {Rare Earth: Why Complex Life Is Uncommon in
  the Universe}.
New York: Copernicus Books

\bibitem[\protect\citeauthoryear{{Zhu} \& {Dong}}{{Zhu} \& {Dong}}{2021}]{ZD21}
{Zhu} W.,  {Dong} S.,  2021, \mn@doi [Annu. Rev. Astron. Astrophys.]
  {10.1146/annurev-astro-112420-020055}, \href
  {https://ui.adsabs.harvard.edu/abs/2021ARA&A..59..291Z} {59, 291}

\makeatother
\end{thebibliography}

\appendix

\section{Probability Density Function: Analytical Calculations}\label{AppA}
In order to determine the posterior probability distribution function, it is necessary to calculate the denominator of (\ref{eq:pdf}). Therefore, the first objective is to determine the following indefinite integral:
\begin{equation}\label{Int}
\mathcal{I}_{\alpha} = \int dp_L\, P(E \vert p_L) P(p_L),
\end{equation}
where $P(p_L)$ for all three cases of interest is represented by the general expression $P(p_L) = \mathcal{C}\, p_L^\alpha$, with $\alpha = 0, -1, -2$ for the optimistic, uninformative, and pessimistic priors, respectively.

In Section \ref{SecBayes}, we have explained why $P(E \vert p_L)$ is given by
\begin{equation}\label{PEpl}
P(E\vert p_L)= 1 - (1-p_L)^{N_H}.
\end{equation}
Thus, on substituting (\ref{PEpl}) and the generic expression for $P(p_L)$ into (\ref{Int}), we eventually arrive at
\begin{eqnarray}\label{Intv2}
\mathcal{I}_{\alpha} &=& \mathcal{C} \int p_L^\alpha \left[1 - (1-p_L)^{N_H}\right] \\ \nonumber
&=& \mathcal{C} \left[\int p_L^\alpha - \int p_L^\alpha (1-p_L)^{N_H}\right] \\ \nonumber
&=& \mathcal{C} \,\frac{p_L^{\alpha+1}}{\alpha+1} \left[1 - {}_2F_{1}\left(\alpha + 1, - N_H; \alpha + 2; p_L\right) \right],
\end{eqnarray}
where the last expression is derived from the properties of the Gaussian hypergeometric function \citep[e.g.,][]{AbSte65}. The scenarios with $\alpha = 0$ (optimistic), $\alpha = -1$ (uninformative), and $\alpha = -2$ (pessimistic) are respectively reduced to
\begin{equation}\label{azeroIndef}
  \mathcal{I}_{0} =  \mathcal{C} \left[p_L + \frac{(1-p_L)^{N_H + 1}}{N_H + 1}\right], 
\end{equation}
\begin{equation}\label{aminusoneIndef}
 \mathcal{I}_{-1} =  \mathcal{C} \left[\ln p_L + B(1-p_L; N_H + 1,0) \right],
\end{equation}
\begin{equation}\label{aminustwoIndef}
\mathcal{I}_{-2} =  \mathcal{C} \left[\frac{(1-p_L)^{N_H + 1}{}_2F_{1}\left(1, N_H; N_H + 2; 1- p_L\right) - (N_H + 1)}{(N_H + 1) p_L} \right],
\end{equation}
where $B(z;a,b)$ is the incomplete beta function in (\ref{aminusoneIndef}). It is worth reiterating we have computed the \emph{indefinite} integrals. The definite integrals are straightforward to calculate once the indefinite integrals and the integration limits are known. The results are furnished below for the three values of $\alpha$ considered herein. The $\alpha = 0$ case is straightforward to determine from (\ref{azeroIndef}), and we obtain
\begin{equation}
\int_0^{p_L} dp_L\, P(E \vert p_L) P(p_L) =  \mathcal{C} \left[p_L + \frac{(1-p_L)^{N_H + 1} - 1}{N_H + 1}\right].
\end{equation}
Evaluating the $\alpha = -1$ scenario, which corresponds to (\ref{aminusoneIndef}), is rendered trickier because of the singularity seemingly manifested at $p_L = 0$, but a careful treatment of this limit leads us to
\begin{multline}
    \int_0^{p_L} dp_L\, P(E \vert p_L) P(p_L) \\ = \mathcal{C} \left[\ln p_L + B(1-p_L; N_H + 1,0) + \Psi(N_H + 1) + \gamma \right],
\end{multline}
where $\Psi(z)$ is the digamma Function and $\gamma$ denotes the well-known Euler–Mascheroni constant \citep[e.g.,][]{AbSte65}. Likewise, the $\alpha = -2$ case associated with (\ref{aminustwoIndef}) is characterised by a similar issue at $p_L = 0$, which proves to be harder to simplify since it is tantamount to an essential singularity; in consequence, we only tackle this function numerically.

Now that the denominator of (\ref{eq:pdf}) is determined, with the salient expressions furnished above, it is straightforward to compute the posterior probability distribution function given by (\ref{eq:pdf}), and the cumulative distribution function thereafter.


\bsp	
\label{lastpage}
\end{document}